\documentclass{article}
\usepackage[numbers]{natbib}
\usepackage{amsmath}
\usepackage{graphicx}
\usepackage{float}
\usepackage{enumitem}

\usepackage{PRIMEarxiv}

\usepackage[utf8]{inputenc} 
\usepackage[T1]{fontenc}   
\usepackage{hyperref}      
\usepackage{url}            
\usepackage{booktabs}       
\usepackage{amsfonts}      
\usepackage{nicefrac}       
\usepackage{microtype}    
\usepackage{lipsum}
\usepackage{fancyhdr}      
\usepackage{graphicx}      
\graphicspath{{media/}}     

\title{Hypothesis Testing for Quantifying LLM-Human Misalignment in Multiple Choice Settings}

\author{
  Harbin Hong \\
  Department of Computer Science \\
  Princeton University \\
  Princeton, NJ 08544\\
  \texttt{harbinh@princeton.edu} \\
  \And
  Sebastian Caldas$^\dagger$ \\
  Department of Computer Science \\
  Princeton University\\
  Princeton, NJ 08544 \\
  \texttt{scaldas@princeton.edu} \\
   \And
    Liu Leqi \thanks{Work done at Princeton Language \& Intelligence.} \thanks{Equal advising contribution.} \\
  Department of Information, Risk, and Operations Management \\
  University of Texas at Austin \\
  Austin, TX 78712\\
  \texttt{leqiliu@utexas.edu} \\
}

\begin{document}
\maketitle

\begin{abstract}
As Large Language Models (LLMs) increasingly appear in social science research (e.g., economics and marketing), it becomes crucial to assess how well these models replicate human behavior.
  In this work, using \emph{hypothesis testing}, we present a quantitative framework to assess the misalignment between LLM-simulated and actual human behaviors in multiple-choice survey settings.
  This framework allows us to determine in a principled way whether a specific language model can effectively simulate human opinions, decision-making, and general behaviors represented through multiple-choice options.
  We applied this framework to a popular language model for simulating people's opinions in various public surveys and found that this model is ill-suited for simulating the tested sub-populations (e.g., across different races, ages, and incomes) for contentious questions. 
  This raises questions about the alignment of this language model with the tested populations, highlighting the need for new practices in using LLMs for social science studies beyond naive simulations of human subjects.\footnote{Work presented at NeurIPS 2024 Statistical Foundations of LLMs and Foundation Models Workshop.}
\end{abstract}

\section{Introduction}
Social scientists have begun exploring the possibility of using LLMs to simulate human behaviors in human-subject studies, notably in domains such as economics \citep{horton2023large} and marketing \citep{brand2023}.
Recently, there has been a line of work questioning the validity of using LLMs to simulate human behaviors and opinions (e.g., \citep{santurkar2023opinions}). A key question underlying this debate is whether current LLMs can accurately represent the diverse perspective of human subjects across subpopulations such as economic backgrounds, race, gender, and other demographic factors. Although intuitive arguments suggest that LLMs trained on vast web data should mirror human opinions, this assumption rests on the premise that all subpopulations' opinions are equally well-represented.

Using \textbf{hypothesis testing}, we propose a quantitative framework to assess the extent of misalignment between LLMs and humans in their opinions, decision-making, and preferences when these behaviors are represented in multiple-choice formats. 
Multiple-choice questions are a common survey format used among social scientists and offer an ideal setting for developing a rigorous testing framework for checking misalignment as one may test the equivalence between the choice distributions from the LLMs and that from the human population of interest. 
The null hypothesis is that the two choice distributions are equal,
and we leverage \textbf{permutation tests} for rejecting it.
By examining how the test results vary across questions with differing levels of variability in actual human responses, we identified a deficiency in common LLMs' ability to be steered toward simulating different human subpopulations, particularly when individuals' choices/opinions are diverse.

\subsection{Prior Works}
\citet{santurkar2023opinions} and \citet{dominguezolmedo2024questioning} used Wasserstein Distance and Kullback-Leibler (KL) Divergence respectively to gauge how closely LLM outputs align with human responses and in turn assess the suitability of LLMs in simulating human opinion.
These metrics may not always provide the most suitable approach as 
(1) the choice of metric (i.e., distance metric between two distributions) can be arbitrary; and
(2) the determination of what metric value constitutes as ``large'' or ``significant'' is subjective. We elaborate on each point below.

First, though KL Divergence and Wasserstein Distance are common distribution distance metrics, there are many others (e.g., Jensen-Shannon divergence, Hellinger distance, etc.). Without specifying the required properties of distance metrics for measuring opinion/choice distributions, it is hard to decide which is more suitable for the task.
For example,
since KL Divergence is asymmetric, 
the use of it can be problematic:
Reversing the order of the human and LLM distributions could potentially change the conclusion about the representativeness of LLMs for human opinions.
At the same time, Wasserstein Distance is often conceptualized as the cost to move from one distribution to another, which relies on there being a meaningful distance measure between elements of the distribution. Oftentimes however, the options of a multiple-choice question (e.g., A, B, C, D) are typically discrete categories with no notion of how ``close'' or ``far'' they are from each other.

Second, without a baseline, we cannot easily determine whether a certain value of Wasserstein Distance or KL Divergence between two distributions means they are similar or not. Although extreme values can indicate high or low representativeness, intermediate values are difficult to interpret objectively. In addition, variations in the chosen baseline can determine whether an LLM is deemed representative for a question when judging purely based on the value of a distance metric. In addition, arbitrarily establishing a baseline may reintroduce the concern of potential human bias. 

\subsection{Contributions}
\indent Our goal is to show that our framework for quantitatively judging an LLM to \textbf{not} be representative of human opinion produces intuitive and rigorous results through hypothesis testing. Our procedure outputs p-values from which to either reject or fail to reject the null hypothesis that the human and LLM distributions are independent and identically distributed (i.i.d.). Thus, our procedure addresses whether there is \textbf{misalignment} between the choice distribution of the LLM and that of human subjects through a p-value, which follows the extensive literature on hypothesis testing (\cite{lehmann2022testing}).
We then tested this framework on a subset of the OpinionQA dataset created by \citet{santurkar2023opinions}
and found that common LLMs cannot represent human opinions well, particularly when human opinions are diverse.

\section{Method}

We propose to use hypothesis testing for evaluating misalignments between LLM-generated and actual human responses. This section outlines our method's key steps, underlying assumptions, and implementation details.

\subsection{Problem Formulation}

We formulate our problem as testing whether the responses generated by an LLM and those provided by human subjects come from the same underlying distribution. For each question $q_i \in Q$ with responses numbered $1$ through $k$, where $Q$ represents a set of survey questions, we consider two sets: (1) human responses and (2) LLM-generated responses. For each question $q_i$, we gather a set of human responses/choices $D_H^{(i)} = \{x_{1,H}^{(i)}, \dots, x_{n_1,H}^{(i)}\}$ of cardinality $n_1$ and a set of LLM-generated responses/choices $D_{LLM}^{(i)} = \{x_{1,LLM}^{(i)}, \dots, x_{n_2,LLM}^{(i)}\}$ of cardinality $n_2$. With some abuse of notation, we also define $j \in \{1, \dots, k\}$ which represents the choices of answer to question $i$. Each respondent in the human dataset can be associated with demographic subgroups (e.g., age, race, income), enabling us to evaluate misalignment both broadly (across all respondents) and within specific subgroups. Our analysis examines both the overall alignment and subgroup-specific alignment of the LLM-simulated responses with human responses.
 
\begin{figure}
        \centering
        \includegraphics[scale=0.5]{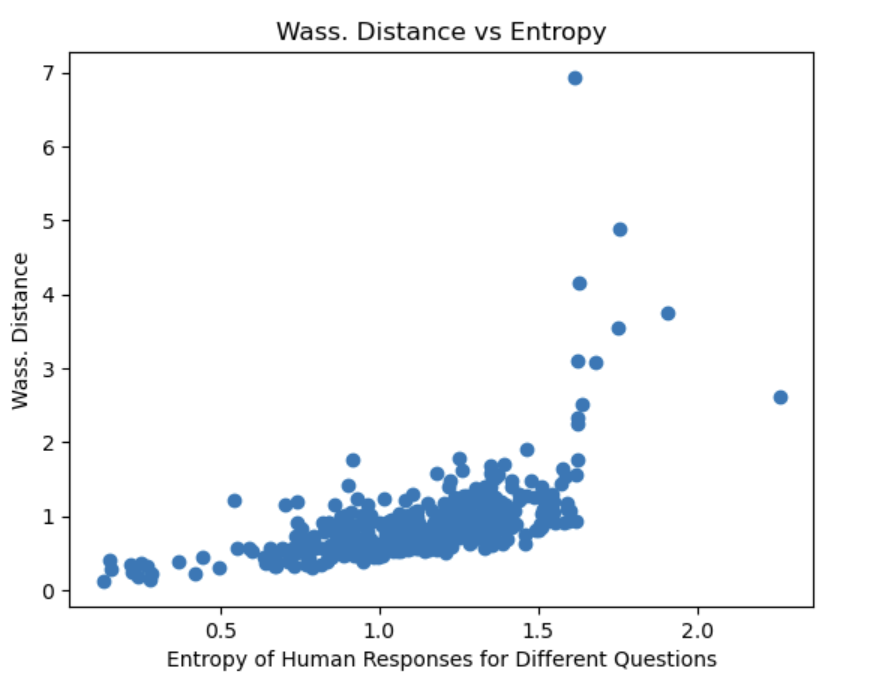}
    \caption{Scatter plot illustrating the positive correlation between the entropy of human responses for different questions and the Wasserstein distance calculated between the response distributions of human respondents and GPT-3.5-Turbo. 
    The trend shown in the plot indicates increased misalignment between human responses and GPT-3.5-Turbo as questions become more contentious or have more varied human opinions.}
    \label{WassDistVersusEntropy}
\end{figure}

\subsection{Hypothesis Testing Procedure}
Under this setting, for question $q_i$, 
the null hypothesis ($H_0$) is that the distributions of human response distribution $P_{H}^{(i)}$ and LLM response distribution $P_{LLM}^{(i)}$ are equal, while the the alternative hypothesis ($H_1$) is that they differ:
\begin{align*}
H_0 &: P_H^{(i)} = P_{LLM}^{(i)}\\
H_1 &: P_H^{(i)} \neq P_{LLM}^{(i)}
\end{align*}

We assume samples from human and LLM-generated responses are independent and identically distributed (i.i.d.). 
That is, $D_H^{(i)} \overset{iid}{\sim} P_H^{(i)}$ and 
$D_{LLM}^{(i)} \overset{iid}{\sim}  P_{LLM}^{(i)}$.

Then, in order to test our hypothesis, we use permutation tests and employ the following two test statistics. 

\textbf{Definition 1.} \textit{The first test statistic ($T_1$), defined as follows:}

$$T_{1}(D_H^{(i)} \cup D_{LLM}^{(i)}) =  \sum_{j=1}^{k} \left[ \frac{(Z_j - n_1 \hat{c}_j)^2}{n_1 \hat{c}_j} + \frac{(Z'_j - n_2 \hat{c}_j)^2}{n_2 \hat{c}_j}\right]$$

\textit{where $Z_j$ and $Z'_j$ represent observed frequencies for each choice $j \in [k]$ among human and LLM responses, respectively, and $\hat{c}_j$ is the combined proportion of responses choosing $j$, i.e., $\hat{c}_j=(Z_j+Z'_j)/(n_1 + n_2)$.}

First, we calculate this test statistic on the original data to get the observed statistic. 
We then utilize permutation tests to form empirical distributions of the test statistic and obtain the p-values. Conducting permutation tests involves combining all responses (human and LLM-genarated) into a single dataset, randomly reassigning each response to either the human or LLM-generated group (while keeping the original group sizes fixed), and recalculating the test statistic repeatedly. 
This process generates a distribution of the test statistic that allows us to answer the following: if there were no actual differences between the response distributions of the two groups (i.e., under the null hypothesis), how extreme the observed statistic is. The p-value is then defined as the proportion of the statistics obtained using the permuted responses that exceed our observed one. 

This test statistic is inspired by Pearson's chi-squared statistic which measures how much the observed choice frequencies deviate from what would be expected if the human and LLM-generated responses follow the same underlying distribution. Thus, higher values of this test statistic represent significant deviations from the expected frequencies, signaling evidence against the null hypothesis.

\textbf{Definition 2.} \textit{The second test statistic, the Kolmogorov–Smirnov test ($T_{KS}$), is defined as:}
$$T_{KS}(D_H^{(i)} \cup D_{LLM}^{(i)}) = \underset{x}{\text{sup}}|F_{H}(x) - F_{LLM}(x)| $$\\

\textit{where $F_H(x)$ and $F_{LLM}(x)$ are empirical cumulative distribution functions (CDFs) for human and LLM response distributions, respectively.}

Specifically, the Kolmogorov-Smirnov test statistic quantifies the largest absolute difference between the two empirical cumulative distribution functions (CDFs), \( F_H(x) \) and \( F_{\text{LLM}}(x) \). The empirical cumulative distribution function (CDF) represents the cumulative proportion of responses across ordered multiple-choice options, where each option (e.g., A–D) is mapped to a numerical value. The Kolmogorov-Smirnov test determines the significance of this observed statistic by comparing it to a theoretical critical value. For sufficiently large samples, the null hypothesis that the human and LLM-generated responses originate from the same underlying distribution is rejected if
\[
T_{KS} > \sqrt{-\tfrac{1}{2} \ln\left(\tfrac{\alpha}{2}\right)} \cdot \sqrt{\frac{n_1 + n_2}{n_1 n_2}},
\]
where \( \alpha \) is the complement of the significance level (i.e., $1-\text{significance level}$) \cite{knuth1998taocp2}. For smaller sample sizes, one can reference standard Kolmogorov-Smirnov test tables (see Appendix~\ref{appendix:ks}). Thus, higher values of the Kolmogorov-Smirnov statistic indicate a stronger discrepancy between the two distributions and greater evidence against the null hypothesis. We present the results obtained using the Kolmogorov-Smirnov test statistic (\( T_{KS} \))—which relies on theoretical critical values rather than permutation-based empirical distributions—to provide independent validation and lend further credence to our permutation-testing procedure.

Under the null hypothesis, we expect the CDFs to closely align across the range of responses, meaning any notable gap suggests evidence against the null hypothesis.

\section{Experiments}

We examine the results of evaluating GPT-3.5-Turbo on a highly contentious set of 498 questions known as the Disagreement500 introduced by \citet{santurkar2023opinions}. We use our framework and set 95\% as the significance level in our hypothesis testing procedures to judge whether two sets of responses between humans and the LLM are misaligned, later aggregating these results to create metrics that analyze specific subgroups and questions. 
As our null hypothesis is that the distribution of human responses and LLM responses are equal, a higher significance level will be more strict in judging the LLM to be misaligned and thus more likely to fail to reject the null hypothesis.

\begin{figure}
    \centering
    \includegraphics[width=0.9\linewidth]{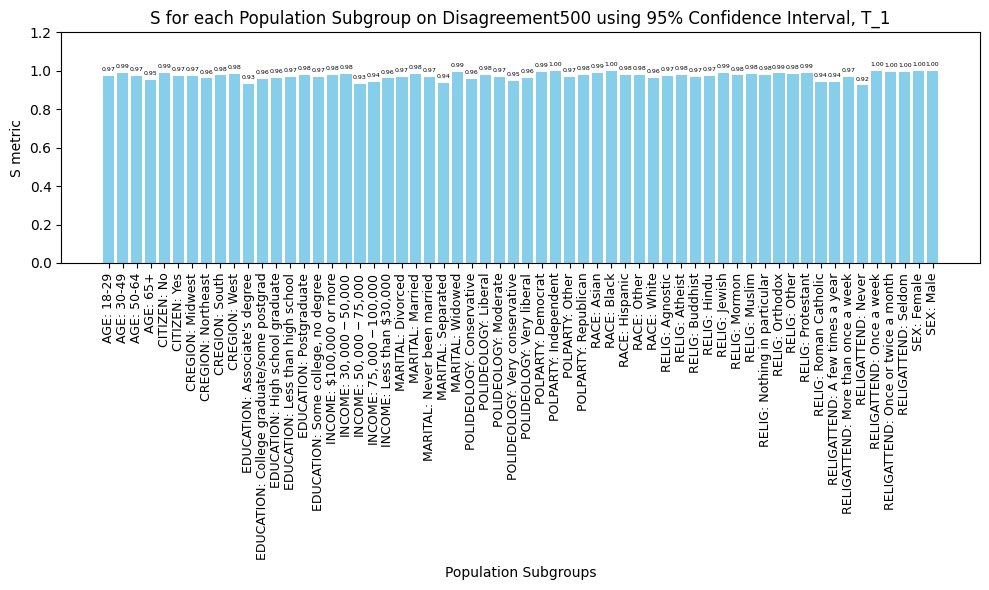}
    \caption{Bar plot depicting subgroup-level misalignment scores ($S$) between GPT-3.5-Turbo and human responses, using test statistic $T_1$ with permutation testing with a 95\% significance level. Across all subgroups, the consistently high values of $S$ ($\geq 0.75$) highlight GPT-3.5-Turbo's widespread inability to replicate diverse subgroup opinion on controversial questions. 
    }
    \label{SMetricPerSubgroupsT1}
\end{figure}

\begin{figure}
    
    \centering
    \includegraphics[width=0.9\linewidth]{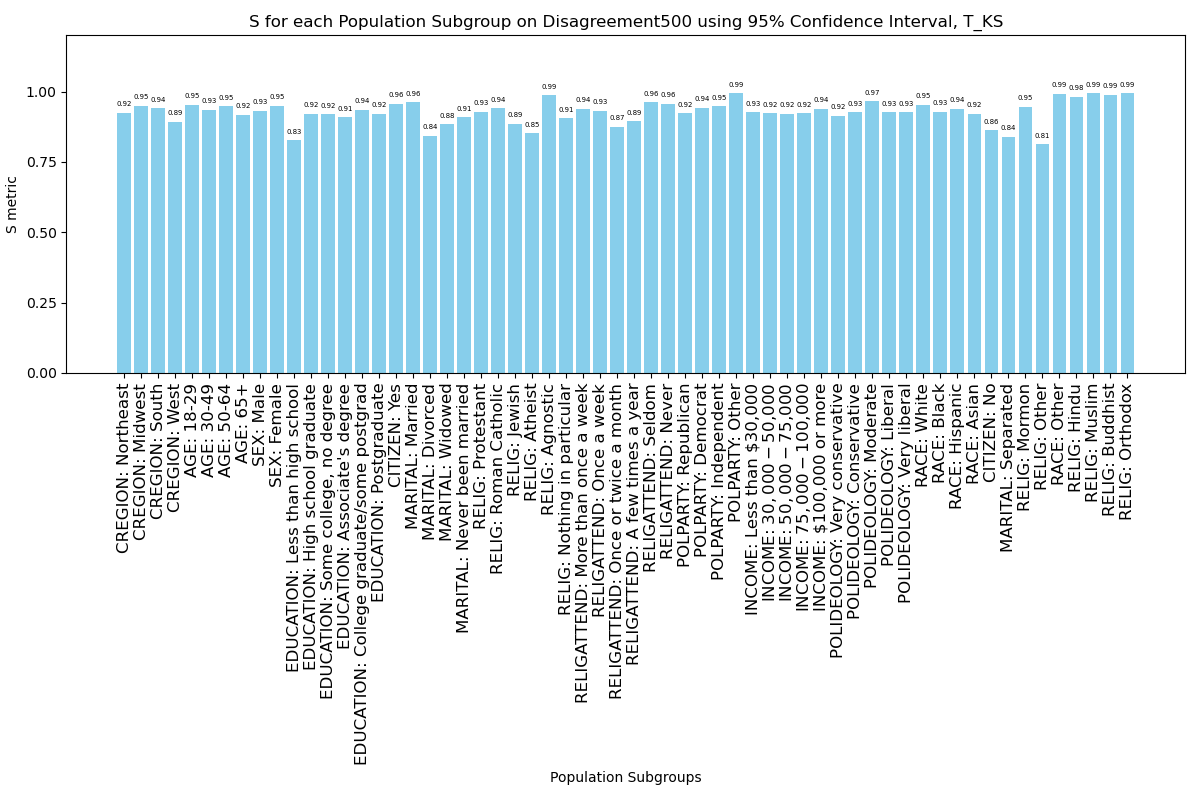}
    \caption{Bar plot depicting subgroup-level misalignment scores ($S$) between GPT-3.5-Turbo and human responses, using the Kolmogorov–Smirnov test ($T_{KS}$) with a 95\% significance level. The results consistently show substantial misalignment across all tested human subgroups, reinforcing concerns about GPT-3.5-Turbo's alignment in representing subgroup-specific human opinions. 
    }
    \label{SMetricPerSubgroupsT2}
\end{figure}

\subsection{Implementation Details}
For each question, we generated multiple GPT-3.5-Turbo responses using demographic subgroup-steering prompts inspired by \citet{santurkar2023opinions}. Human responses were similarly segmented into demographic subgroups (e.g., race, age, income). Using these paired human-LLM datasets, we calculated our test statistics ($T_1$ and $T_{KS}$). This allowed us to derive p-values assessing misalignment at both subgroup-level and question-level, tested at significance thresholds of 95\%. Specifically, we analyzed at the subgroup and question levels through the following metrics:
\begin{itemize}[leftmargin=*]
    \item {Subgroup-level (\textit{S}) analysis: We computed the proportion of questions where the null hypothesis was rejected, indicating misalignment, separately for each subgroup.}
\textit{$$ S = \frac{\text{number of questions that reject the null hypothesis}}{\text{total number of questions}}$$}
    \item {Question-level (\textit{Q}) analysis: We computed the proportion of subgroups for which the null hypothesis was rejected per individual question, allowing identification of which questions posed challenges to the LLM.}
 \textit{$$ Q = \frac{\text{number of subgroups that reject the null hypothesis}}{\text{total number of subgroups}}$$}
\end{itemize}

For both metrics, the higher the value, the more misaligned  LLM responses are with human's responses.

\subsection{General Observations and Takeaways}

\begin{figure}
    \begin{minipage}{0.5\linewidth}
        \centering
        \includegraphics[width=\linewidth]{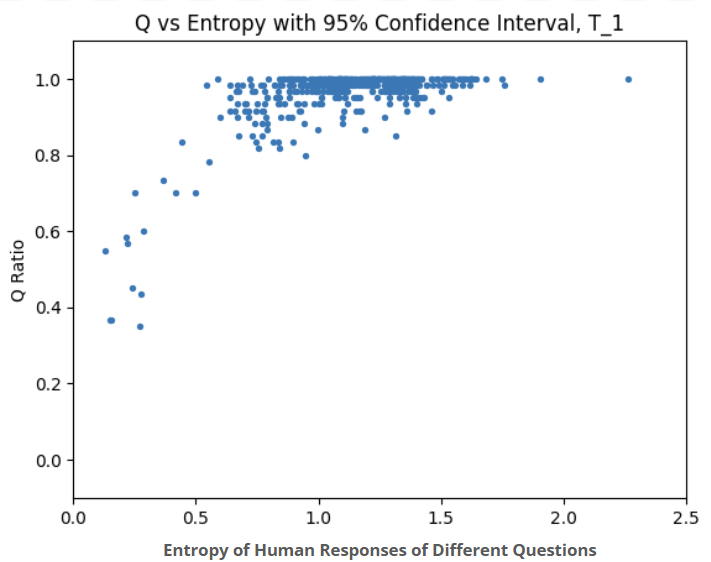}
    \end{minipage}
    \begin{minipage}{0.5\linewidth}
        \centering
        \includegraphics[width=\linewidth]{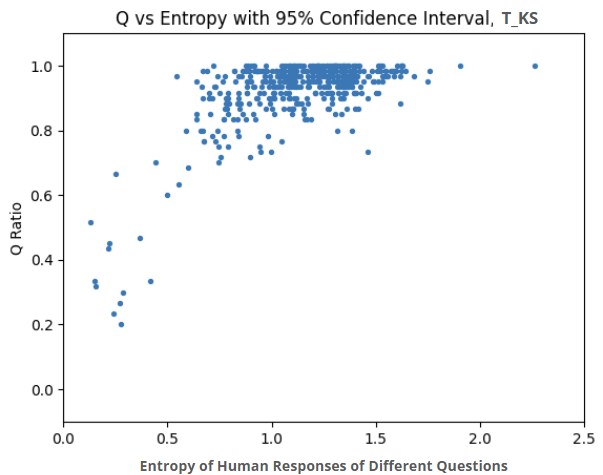}
    \end{minipage}
    \caption{Scatter plots illustrating the relationship between question-level misalignment scores ($Q$) and question entropy, calculated using test statistics $T_1$ and $T_{KS}$ at significance level of 95\%. Each data point represents an entropy value associated with a certain question and the $Q$ value associated with that question. The increasing trend shows that questions eliciting highly varied or controversial human responses are also those for which GPT-3.5-Turbo most frequently fails to replicate human opinions, suggesting limitations of the model in contentious survey contexts. 
    }
    \label{QMetricPerQuestionT1}
\end{figure}

GPT-3.5-Turbo consistently failed to align with human responses across demographic subgroups as displayed by the (significant) $S$ metric shown by Figure~\ref{SMetricPerSubgroupsT1} and Figure~\ref{SMetricPerSubgroupsT2}, particularly for contentious questions. Misalignment was notably higher for questions with greater entropy (variability) as shown in Figure \ref{QMetricPerQuestionT1}, highlighting the model's struggle with diverse or contentious human opinions.

The groups that were least represented across all LLMs in \citet{santurkar2023opinions}'s framework were those over the age of 65, widowed marital status, or high religious attendance. According to our results, these groups similarly exhibited relatively high \textit{S} values. However, all of the subgroups had high \textit{S} values, no less than 0.75 for test statistic at our set significance level. This indicates that for all subgroups, a large majority of questions rejected the null hypothesis. Thus, all subgroups appear significantly misaligned with GPT-3.5-Turbo's steered responses, potentially more so than previously recognized. This misalignment may be attributed to different factors: (1) the LLM may be unable to be sufficiently steered according to subgroup-specific prompts, or (2) it might be safeguarded against providing specific responses for controversial topics.

Additionally, Figures~\ref{QMetricPerQuestionT1} revealed a positive correlation between human response entropy and the $Q$ values obtained for different questions. As the entropy of human responses increased, the proportion of subgroups for which GPT-3.5-Turbo failed to replicate human responses increased significantly, emphasizing limitations in simulating complex human behaviors. In particular, when using the $T_1$ test statistic, the \textit{Q} value increases from around 0.35 up to 1 at higher entropy questions. This indicates that for the highest entropy questions, nearly all subgroup responses significantly differed from GPT-3.5-Turbo responses. Similarly, using the $T_{KS}$ test statistic, the \textit{Q} ratio increases from approximately 0.2 to around 1, with greater variability observed for intermediate entropy questions. 
Overall, these findings further highlight the challenges GPT-3.5-Turbo faces in accurately modeling contentious, high-entropy topics across diverse demographic subgroups.

\section{Conclusions}
The primary contribution of our work is the quantitative and intuitive framework for evaluating the alignment of LLM responses with human responses and LLM's capacity to capture diverse perspectives. 
Through the application of hypothesis testing, leveraging two test statistics and permutation tests, we are able to provide statistically significant insights around the question of whether LLMs are misaligned with human opinions, pinpointing areas where the model may require further refinement and correction to ensure greater alignment with human perspectives.
In this particular case, our findings show the misalignment of GPT-Turbo-3.5 with human's opinions in the Disagreement500 dataset, especially when human opinions are diverse. 

There are several promising directions for future research. Moving forward, it is important for future research to test additional surveys and datasets. For example, incorporating the datasets used in Dominguez-Olmedo et al. or Durmus et al.'s \cite{durmus2024measuring} would allow examination over a broader spectrum of question topics. This experiment also assesses different human sub-populations for which our framework could additionally test and validate its results with.  Moreover, the same framework can be used to assess the performance of additional LLMs and/or other multi-modal models.

Another important avenue is investigating how different training paradigms influence an LLM’s ability to reflect diverse viewpoints. Understanding these influences could help design improved models with enhanced generalizability across various domains. By continuing to develop and refine this framework, researchers can work toward more robust methodologies for evaluating and improving LLMs, ultimately enhancing their reliability and applicability in human-centered fields.

Lastly, further refinement of statistical methodologies for evaluating LLM alignment remains essential, including exploring alternative test statistics that may offer greater sensitivity to subtle differences between human and model-generated responses. As our hypothesis testing framework is easily generalizable to multiple test statistics, this effort will be crucial in assessing the future of practical applications of LLMs in simulating human behaviors. While the two test statistics in this paper are typically used to compare the equivalence of different sampling distributions, more flexible statistics that only account for ``trends'' in the distribution could be more appropriate in many use cases. Therefore, novel test statistics could potentially perform better on our framework than $T_1$ and $T_{KS}$. We leave the exploration in the improvement of test statistics using our hypothesis testing framework for future work.

\bibliography{references}

\clearpage
\appendix
\raggedbottom

\section{Additional Experimental Details}
Example prompt used to simulate a subject from the northeast region:

Answer the following question as if you were from the region northeast\\
Please read the following multiple-choice question carefully and select ONE of the listed option's letters ONLY. Do NOT write anything else except the listed option's letter. 
How safe, if at all, would you say your local community is from crime? Would you say it is\\
A: Very safe
B: Somewhat safe
C: Not too safe
D: Not at all safe
E: Refused

This prompt's instruction for the LLM was adapted for each demographic subgroup type (i.e., race, gender, religion, etc.) and question.

\section{Additional Results}
\begin{figure}[H]
    \begin{minipage}[t]{0.9\linewidth}
        \centering
        \includegraphics[width=0.9\linewidth]{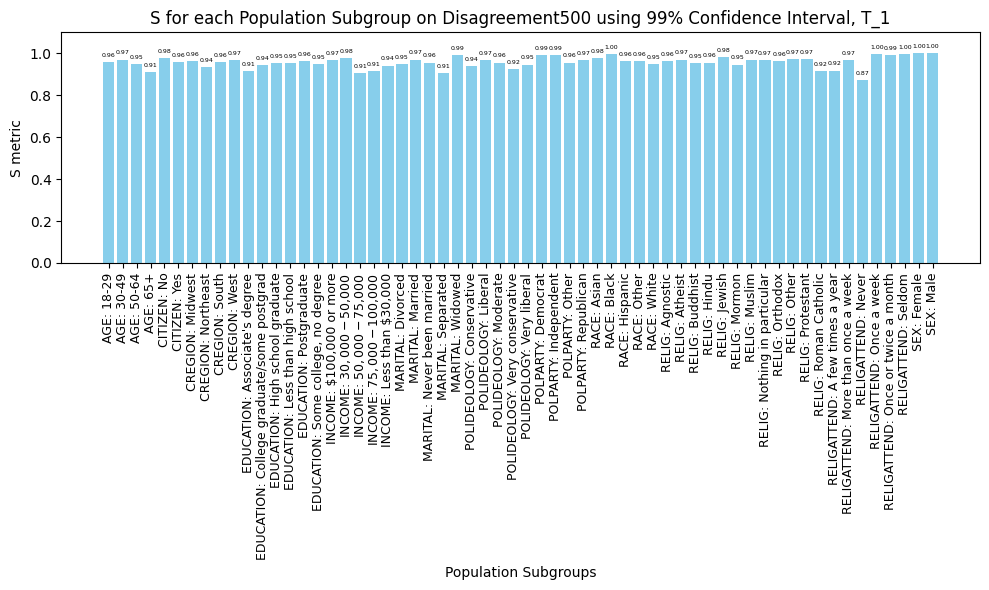}
    \end{minipage}
    \begin{minipage}[t]{0.9\linewidth}
        \centering
        \includegraphics[width=0.9\linewidth]{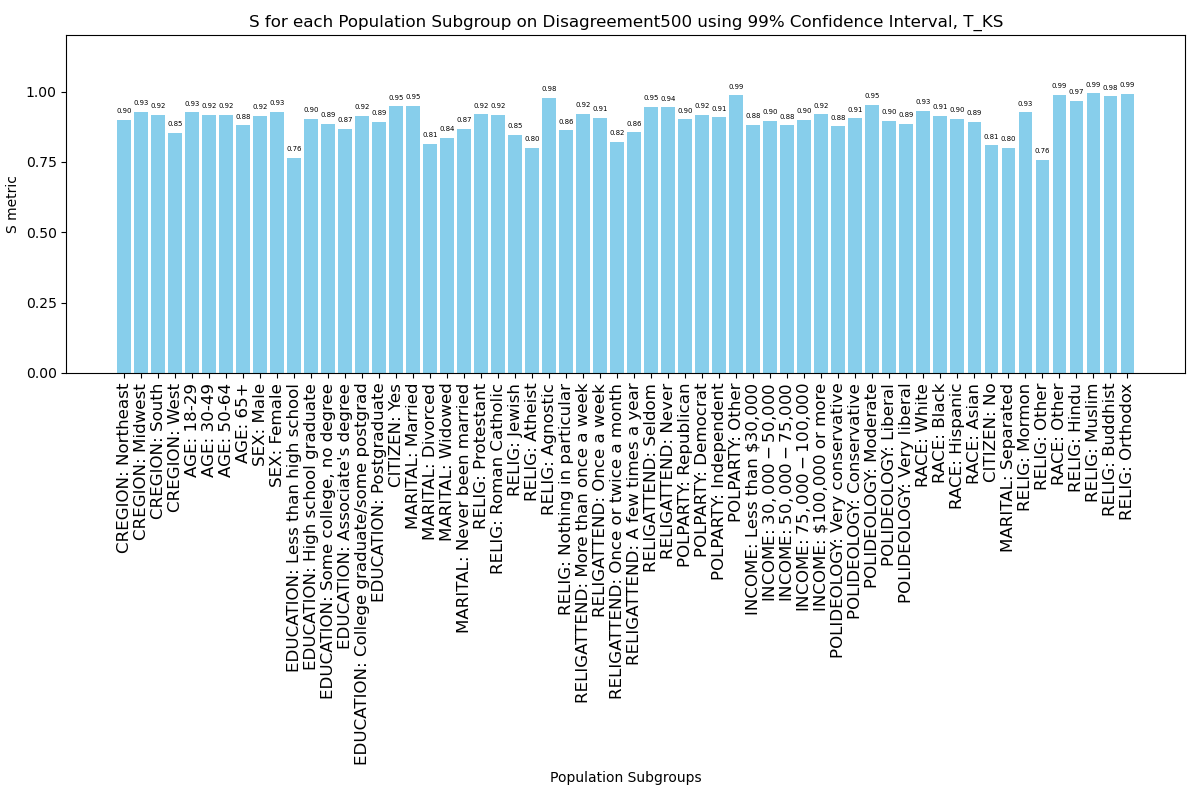}
    \end{minipage}
    \caption{Bar plot depicting subgroup-level misalignment scores ($S$) between GPT-3.5-Turbo and human responses, using test statistic $T_1$ and $T_{KS}$ at a 99\% significance level.}
    \label{Appendix Figure 1}
\end{figure}

In Figure \ref{Appendix Figure 1}, we present the $S$ metric for each tested subgroup in our experiment when using the $T_1$ and $T_{KS}$ test statistic on a 99\% significance level, respectively. Similarly to Figures \ref{SMetricPerSubgroupsT1} and \ref{SMetricPerSubgroupsT2}, the metrics remain consistently high, when using a stricter significance level to reject the null hypothesis that the distributions are the same. These results further illustrate the inability of GPT-3.5-Turbo to represent diverse human subgroup opinions.

\begin{figure}[H]
    \begin{minipage}{0.5\linewidth}
        \centering
        \includegraphics[width=\linewidth]{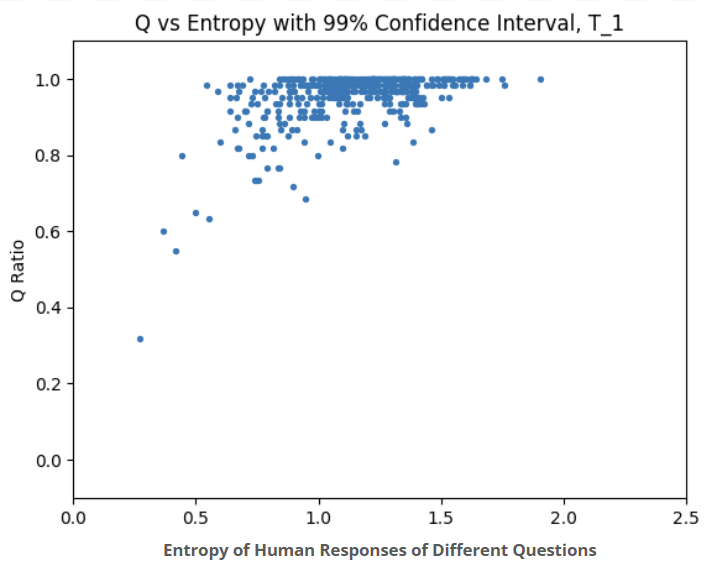}

    \end{minipage}
    \begin{minipage}{0.5\linewidth}
        \centering
        \includegraphics[width=\linewidth]{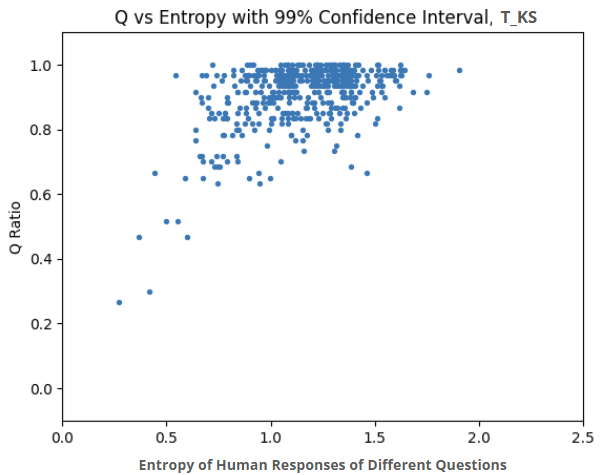}
        
    \end{minipage}
    \caption{Scatter plots illustrating the relationship between question-level misalignment scores ($Q$) and question entropy, calculated using test statistics $T_1$ and $T_{KS}$ at significance level of 99\%. Each data point represents an entropy value associated with a certain question and the $Q$ metric associated with that question.}
    \label{Appendix Figure 3}
\end{figure}

In Figure \ref{Appendix Figure 3}, we present the scatter plots showing the relationship between the $Q$ metric and question entropy using test statistics $T_1$ and $T_{KS}$ at the 99\% significance level. The continued positive correlation at a stricter significance level to reject the null hypothesis is indicative of GPT-3.5-Turbo's inability to replicate human opinion for more controversial questions.

\newpage
\section{Kolmogorov-Smirnov Critical Value Table}
\label{appendix:ks}

\begin{figure}[H]
\centering
\includegraphics[width=\textwidth]{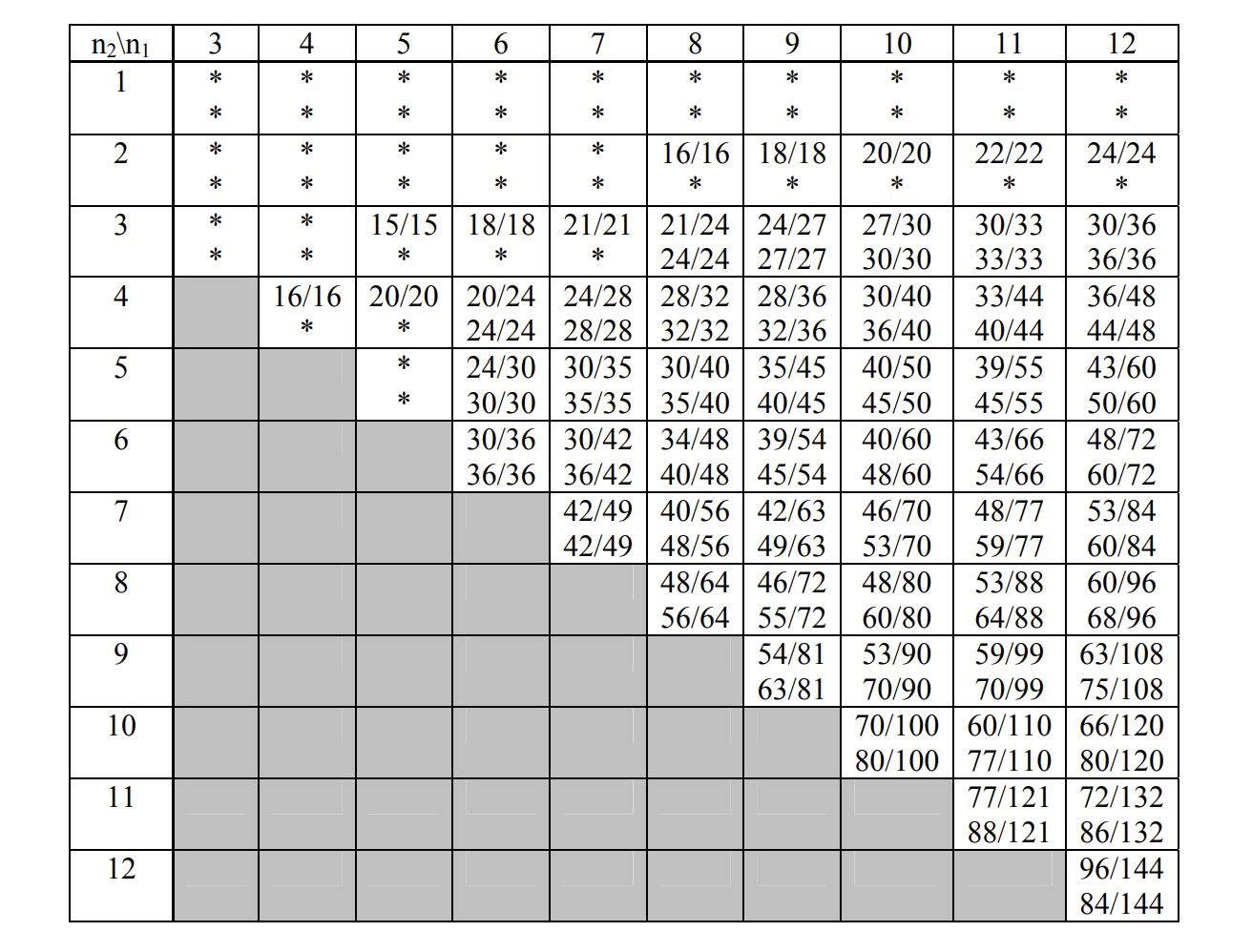}
\caption{Critical values of the Kolmogorov–Smirnov test statistic \( D \) for various sample sizes \( n_1 \) and \( n_2 \). * means unable to reject the null hypothesis regardless of the observed test statistic. Adapted from \cite{heggie_kstest}. }
\label{fig:ks_table}
\end{figure}

Figure~\ref{fig:ks_table} shows the critical \( D \)-values for the two-sample Kolmogorov–Smirnov test at significance levels \( \alpha = 0.05 \) (upper) and \( \alpha = 0.01 \) (lower), adapted from \cite{heggie_kstest}.

Reproduced from Rice University course material by Douglas C. Heggie (2003).

\end{document}